\documentclass[a4paper,11pt]{article}
\pdfoutput=1 

\usepackage{jinstpub} 
\usepackage{url}
\usepackage{comment}
\usepackage{subfig}
\usepackage{tabularx}
\usepackage{graphicx}
\usepackage{amsmath}
\usepackage{amsfonts}
\usepackage{amssymb}

\usepackage{soul}

\title{Shower Identification in Calorimeter using Deep Learning}

\author[a,1]{Y. Verma,\note{Corresponding author.}}
\author[a,2]{S. Jena\note{Corresponding author.}}

\affiliation[a]{Indian Institute of Science Education and Research,\\Sector-81, Knowledge City, SAS Nagar, Punjab, India}

\emailAdd{ms16027@iisermohali.ac.in}
\emailAdd{sjena@iisermohali.ac.in}

\abstract{ 
Pions constitute nearly $70\%$ of final state particles in ultra high energy collisions. They act as a probe to understand the statistical properties of Quantum Chromodynamics (QCD) matter i.e. Quark Gluon Plasma (QGP) created in such relativistic heavy ion collisions (HIC). Apart from this, direct photons are  the most versatile tools to study relativistic HIC. They are produced, by various mechanisms, during the entire space-time history of the strongly interacting system. Direct photons provide measure of jet-quenching when compared with other quark or gluon jets. The $\pi^{0}$ decay into two photons make the identification of non-correlated gamma coming from another process cumbersome in the Electromagnetic Calorimeter. We investigate the use of deep learning architecture for reconstruction and identification of single as well as multi particles showers produced in calorimeter by particles created in high energy collisions. We utilize the data of electromagnetic shower at calorimeter cell-level to train the network  and show improvements for identification and characterization. These networks are fast and computationally inexpensive for particle shower identification and reconstruction for current and future experiments at particle colliders.

}

\begin{document}
\maketitle
\flushbottom

\section{Introduction}
\label{sec:intro}

Detector acts as an imaging equipment for the final state particles produced in high energy collisions. Many types of detectors ranging from Gas detectors such as Multigap Resistive Plate Chamber (MRPC)~\cite{mrpc}, Gas Electron Multiplier (GEM), MicroMegas etc to Scintillators based detectors are being used to identify charged as well neutral particles produced in collision events. In experiments, numerous detectors are arranged in specific sequence to probe the properties of particles produced in collision experiment. For example, in CMS experiment~\cite{cms}, various layers of detectors are arranged and each of them measures the tracks. The first detector is the tracker detector which records the paths taken by charged particles. Then there is Electromagnetic calorimeter (ECAL) which detect particles that interact mainly through electromagnetic interaction like $\gamma,e$ and measures the energy and their shower profiles. Similar to ECAL, Hadronic Calorimeter (HCAL) is placed which detect particles that interact through strong interaction like proton, neutron etc.  The last layer of detectors consist of muon detector as muons are highly penetrating particles and are not stopped by calorimeters. Other experiments like ALICE~\cite{alice} ,ATLAS~\cite{atlas} etc follows a similar approach of detector sequence.

Among all these detectors, calorimeters act as an imaging equipment for the energy of the incoming particle. When a particle traverses through the calorimeter, a shower of secondary particles produces depending on the type of the calorimeter; for ECAL electrons-positrons pairs are produced whereas for HCAL hadron showers are produced. These traversed particles leave their footprint in the detector in terms of energy depositions in the calorimeter. The energy deposited in the calorimeter is used to perform various task like reconstruction of energy of the incoming particle, type of particle etc. The calorimeter volume can be segmented into smaller volumes called unit cells. Then the energy information of each unit cell can be utilized to form a set of pixelized images which are sensitive to the type and energy of the particle.

Before discussing about detailed techniques of particle reconstruction in the detector, it is essential to know about the various particle types and  their production mechanisms in HIC. Among all the particles produce, pions constitute a high fraction of final state particles in heavy ion collisions and are extensively studied to probe the statistical features of QGP created in collision~\cite{pt,pion}. Majority of the $\pi^{0}$ decays into two $\gamma$ becoming a major source of correlated $\gamma$ production whereas, a direct photon is a photon which originates directly from an electromagnetic vertex in a quark-quark, quark-gluon or gluon-gluon scattering sub-process or originate from compton and annihilation processes. Direct photon being an electromagnetic particle, doesn't interact with strongly interacting medium, hence carries early information about the system produced in heavy ion collision. These direct photons acts as a probe to study jet quenching effects, viscous dynamics of the QGP etc. Although, direct photons are very good messenger of the system, but its detection becomes extremely challenging because of the other source of $\gamma$ coming from various meson decays.

Therefore, it is necessary to distinguish between showers of the correlated $\gamma$ coming from neutral pion ($\pi^{0}$) decays and any random $\gamma$. One of the challenging task is to resolve the showers of $\pi^{0}$ and $\gamma$ in the calorimeter. This is due to the reason that the decay angle of the two $\gamma$ emitted from $\pi^{0}$ decreases with the increase in the energy of the $\pi^{0}$. This led to the two $\gamma$ becoming more and more collimated with the increase in energy of parent $\pi^{0}$. Thus, the shower originated from the collimated  $\gamma$ are hard to distinguish from a direct $\gamma$. Moreover, a simultaneous shower originating from multi-particles states like $\pi^{0}\gamma$, $\gamma\gamma$ will increase the complexity to resolve the showers and identify the originating particles.

Traditionally, several algorithms are used to perform these tasks producing  various measurements  of  physical features like shower width, reconstructed energy, rate of energy loss for particles traversing calorimeter layers and identification of particles ~\cite{reconst,reconst2,reconst3,reconst4}. When we move to high luminosity experiments, like HL-LHC~\cite{hllhc}, the sensitivity of the detector is required to be improved not only to precisely measure the $\gamma$ but also to be able to isolate rare phenomenon. The accelerator upgrades will lead to increase in data volumes and pose a various technological and computational challenges in tasks ranging from tracking to reconstruction and identification. When moving to HL-LHC~\cite{hllhc} and other future accelerator upgrades will lead to higher data volume pose numerous challenges both technologically and computationally. These challenges require novel methods which has led to application of advanced computing techniques in calorimetry. Although, traditional techniques remain very successful to reconstruct the data, however the large data volume at high luminosity becomes a huge bottleneck for these algorithm to work. Therefore, the advanced techniques like machine learning algorithm can be adapted to overcome these issues. The application of machine learning (ML) in HEP have seen an exponential rise from niche field~\cite{ml1,ml2,ml3} to wider applications~\cite{ml4,ml5,ml6,ml7,ml8}. In traditional algorithms, mostly shower topologies are not considered whereas these shower topologies may become an interesting feature for identification and reconstruction tasks. These extra features can be used as an input to machine learning algorithms to preform these tasks. Neural networks have been proved to be very efficient among several machine learning techniques in handling numerous features to execute various identification and reconstruction tasks.

In this article, we demonstrate the application of neural network based approach to identification of particle and multi-particle states from 3-D shower profile in electromagnetic calorimeter. To do this, we used fully simulated calorimeter data by modelling a 3D $Si$ - $PbWO_4$ electromagnetic calorimeter using GEANT4~\cite{geant} . We developed a ML model, deep neural network,  that receive 3D shower profile from calorimeter data as form of  calorimeter-cell voxelized images from single particle as well multi-particle showers and identify the particle/multi-particle states corresponding to that shower.  We limit our model to take data from Electromagnetic calorimeter (ECAL) showers to demonstrate the effectiveness of the model to identify particles states relying only upon the one type of dataset. The deep network models were implemented and trained using Keras~\cite{keras}  and  Tensorflow ~\cite{tensorflow}. In  section~\ref{sec:detector}, we describe the schematics of the detector modelled using GEANT4. Section~\ref{sec:data} depicts the representation of dataset generated for $\pi^{0}$ and $\gamma$. Sections ~\ref{sec:process} and ~\ref{sec:dnn} describe the pre-processing steps we follow and description of the deep network model we considered. In particular, section~\ref{sec:result} shows results of our model in desired task of identification of showers in detector. Conclusions are given in section~\ref{sec:conc}.

\section{Design of $Si$ - $PbWO_4$ Calorimeter}
\label{sec:detector}
The main goal is to detect various particles by optimizing the design of calorimeter with  adequate  energy  and position resolutions which can handle extreme high particle density environment. We constructed a hypothetical electromagnetic calorimeter to carry out our studies. A  sampling  type  electromagnetic  calorimeter of dimension $10cm\times10cm\times22cm$ with silicon cells as sensitive medium and lead tungstate ($PbWO_4$) as an absorber has been modeled using GEANT4 simulation toolkit~\cite{geant}.  The detector consist of 20 layers of lead tungstate and 20 layers of $Si$ pad-arrays layer. Each absorber layer is of dimension $10cm\times10cm\times0.5cm$ followed by a Si pad-array layer. We have chosed the lead tungstate due to the fact that it is a high-Z element scintillator which converts the high energy $\gamma$ and electrons into electromagnetic showers. 

\begin{figure}[!hbt]
        \centering
        \includegraphics[scale=0.4]{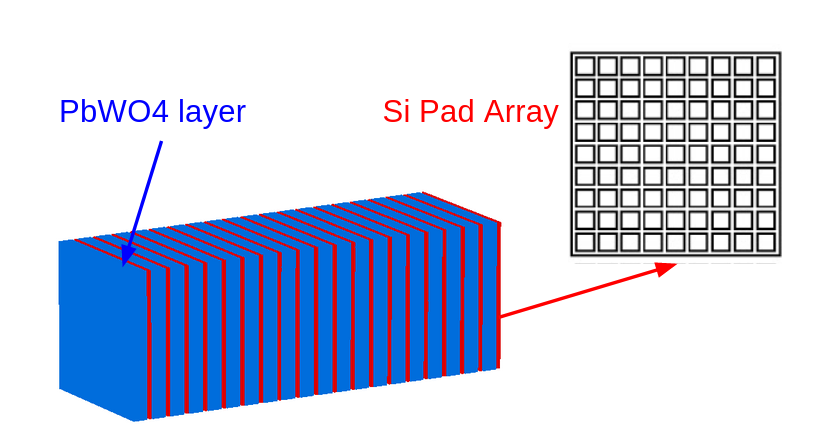}
        \caption{(color online) Schematic view of $Si$ - $PbWO_4$ ECAL}
        \label{fig:det}
\end{figure}

One of the major goal is to resolve the showers associated from $\pi^{0}$ mesons and other $\gamma$. The major complexity comes from the fact that the opening angle of emitted $\gamma$ from $\pi^{0}$ decreases with the increase in energy, mimicking it as if it is a single $\gamma$. Therefore, our first task is to find out the $\gamma$ and photons originated from $\pi^0$ with respect to its incoming energy. Thus, a detail knowledge of the properties of each of them is required.  In order to probe the properties of incoming particle, tracking and good position resolution of showers in sensitive layer is essential. The sensitive layer, active sensor layers, in calorimeters are built by considering all of these facts. Twenty sensitive layer of high granularity are considered which consist of 900 silicon cells in each layer, where each cell is of $0.3cm\times0.3cm\times1.5mm$ dimension. A pictorial representation of detector is shown in Fig.~\ref{fig:det}.

\section{Dataset}
\label{sec:data}

\begin{figure}[!hbt]
        \centering
        \includegraphics[scale=0.45]{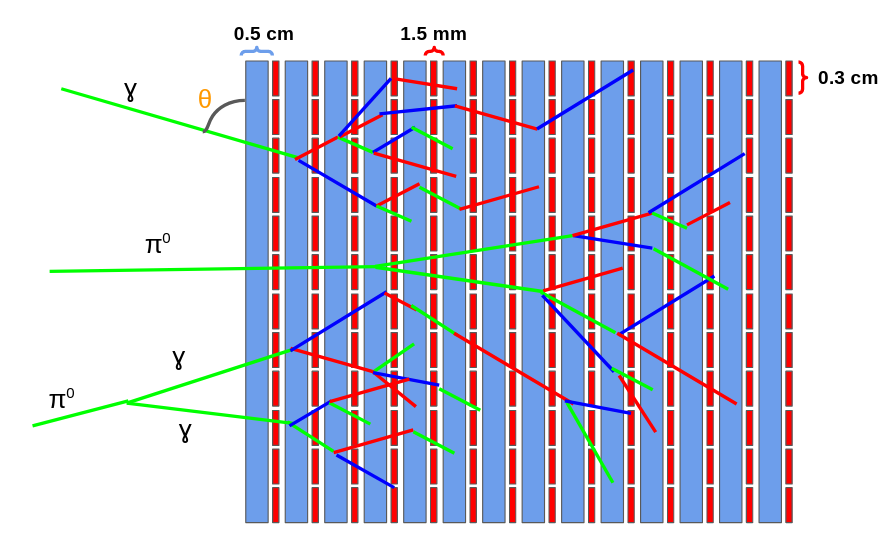}
        \caption{(color online) Pictorial representation of shower profile in detector for $\gamma$ and $\pi^{0}$ in first few layers with $\theta$ is the incident angle}
        \label{fig:shodet}
\end{figure}

\begin{figure}[!hbt]
\def\tabularxcolumn#1{m{#1}}
\begin{tabularx}{\linewidth}{@{}cXX@{}}
\begin{tabular}{cc}
\subfloat[$\gamma$ Shower]{\hspace{-5.0em}\includegraphics[scale=0.7]{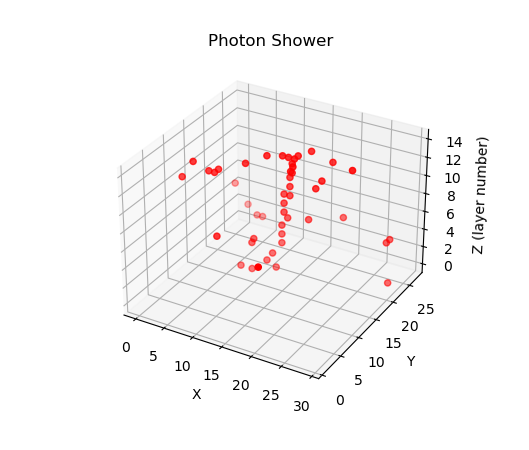}} 
   & \subfloat[$\pi^{0}$ shower]{\hspace{-2.0em}\includegraphics[scale=0.75]{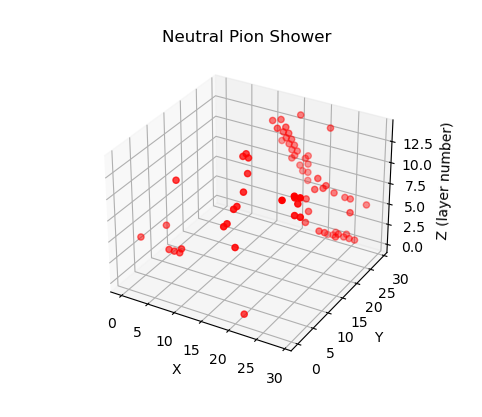}}\\
\end{tabular}
\end{tabularx}
\caption{3D image of a photon (left) and neutral pion (right) shower in detector for one event at $1~GeV$ energy. The red points are hit-points in the Silicon pad arrays. }\label{fig:detect}
\end{figure}

To train the neural network model to identify shower signatures of $\gamma$ and $\pi^{0}$ and its combinations in the detector, we first simulated $\gamma$ and $\pi^{0}$ shower in the $Si-PbWO_4$ calorimeter using GEANT4 for various energies ranging from $100~MeV$ to $5~GeV$ in random position in x-y plane and projecting towards front end of detector. Particles are simulated by throwing at random angles to the face plane of the detector to maximally mimic experimental scenario. A schematic view of shower profile as hit-points in Si pad array is shown in Fig.~\ref{fig:detect}, \ref{fig:shodet} for both $\gamma$ and $\pi^{0}$ for an event. 

In the experiment scenario, it is also possible that simultaneous shower from two or three  $\gamma$ and $\pi^{0}$  can traverse through detector at same instant creating shower. Hence, it is essential to characterize and identify these showers coming from multiple particles. So, we also simulated the events corresponding to showers originating from multi-particles like $\gamma\gamma$, $\pi^{0}\gamma$ etc.  

\newpage
\section{Deep Network Model}
\label{sec:dnn}

We implemented the 3D Convolutional Neural Network (CNN) model~\cite{cnn} for end to end reconstruction and classification of showers of different particles in the detector. The CNN network consist of four 3D convolutional layers that act on the input 3D image. The first layer (conv1) consist of $64$ filters having a size of $(2,2,2)$, the second layer(conv2) of $32$ filters having same size. The third (conv3) and fourth (conv4) layer consist of $16$ filters having the same size. After each convolutional layer, dropout is also implemented with a rate of $0.5$ to curb over-fitting of the model. Batch Normalization is used after second and third layer in addition to the use of 3D Max Pooling having size of $(2,2,2)$. 

\begin{figure}[!hbt]
        \centering
        \includegraphics[scale=0.5]{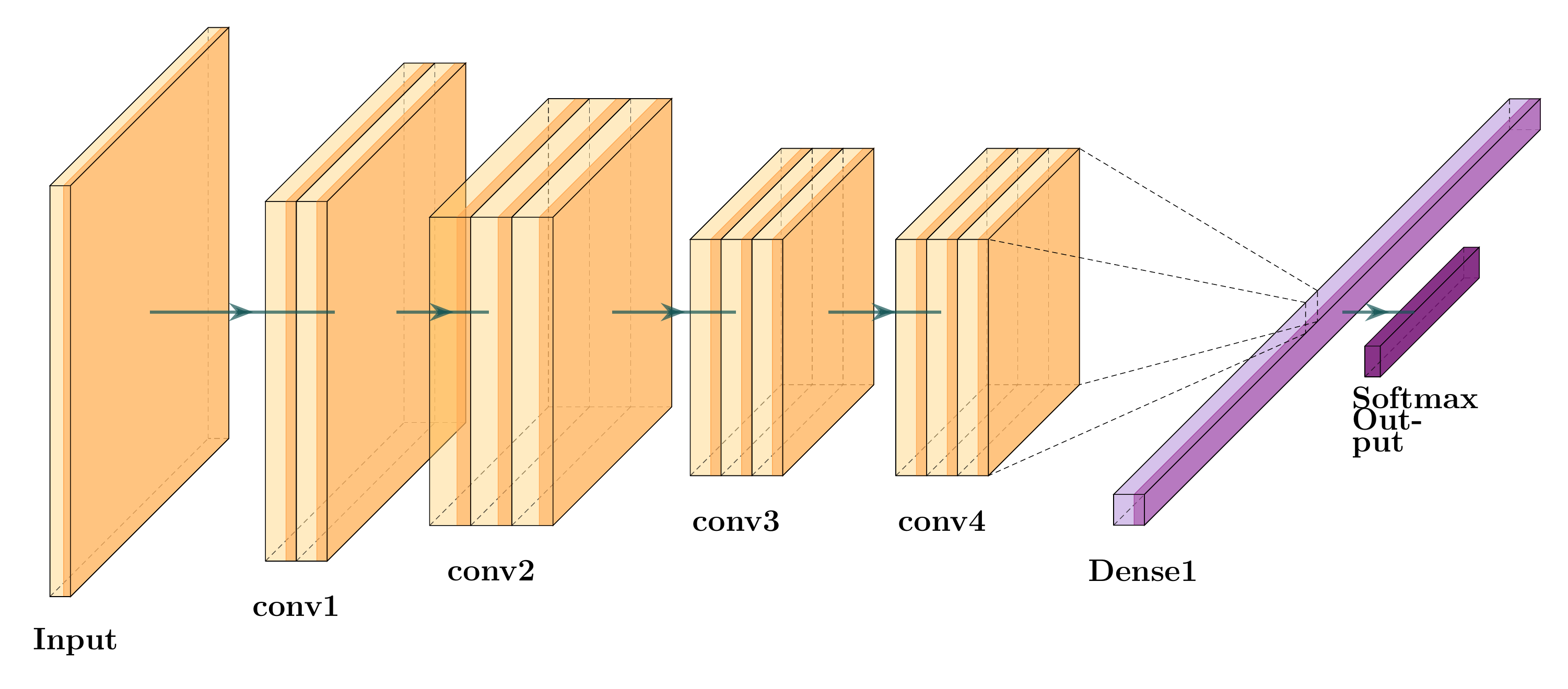}
        \caption{(color online) Schematic view of model architecture }
        \label{fig:cnn}
\end{figure}

The convolution network layer is flattened after fourth layer to Dense network. A dense layer (Dense1) of 100 neurons and a final layer of two neurons with \emph{softmax} activation function which output the probability of event belonging to various classes. \emph{Sigmoid} activation function is implemented in all the layers except the last dense layer of two neurons. A schematic view of the model architecture is shown in Fig.~\ref{fig:cnn}

\newpage
\section{Dataset Preprocessing}
\label{sec:process}

The input to our model for training is a volume of images $\rho (E,x,y,z)$ representing the energy deposition in the silicon cells in an event. It is a three-dimensional energy signature representation of the shower pattern of constituent particles. The  image consists of a 3D regular grid of pixels in $x \times y 
\times z$ giving us a matrix of $30 \times 30 \times 20$ which is $\rho$.

The intensity of each cell or each entry in the above matrix is the $E (x,y,z)$ (the energy deposited in the respective cell during an event). Opposite to the commonly studied data-sets of images, these images are highly sparse and do not posses smooth features. This serves as an input to our model for training and testing. The class or output labels are converted to one hot encoding vectors by converting them into categorical variables for training and testing. We could not include a very large area detector due to limitations in computational resources.

\section{Results}
\label{sec:result}

\begin{figure}[h!]
    \centering
    \includegraphics[scale=0.6]{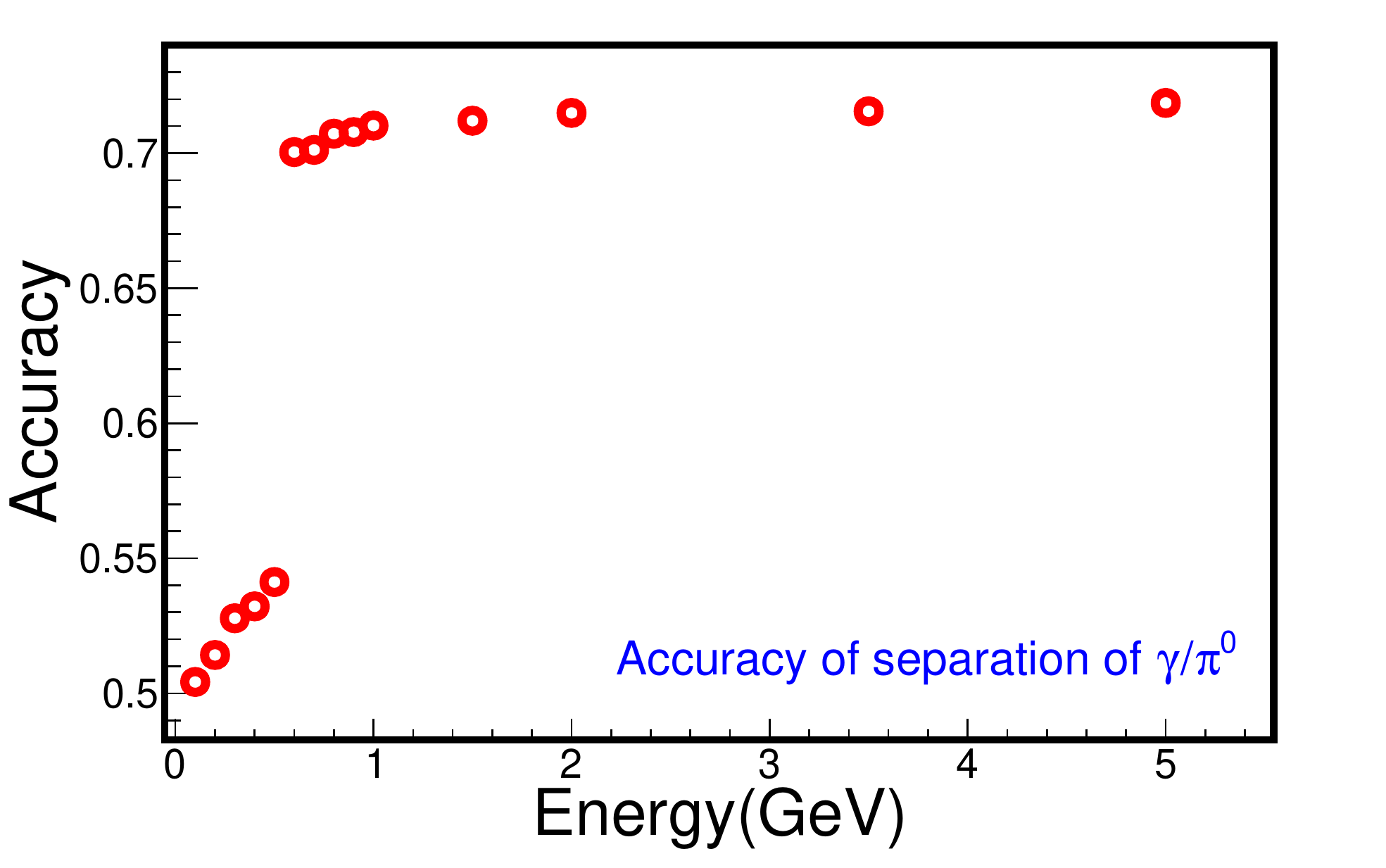}
    \caption{Accuracy of classification between $\gamma/\pi^{0}$ by 3D ConvNet model  }
    \label{fig:single}
\end{figure}

The model was trained on 100k samples for each event and validation was done on 30k samples of each event. We trained a binary classifier using our deep network model to classify between various shower profiles as a function of their energy signature in the detector for various energies. We first start from basic task of classifying $\gamma$ vs $\pi^{0}$. The validation accuracy for the model tested after the model has been trained is shown in Fig.~\ref{fig:single}. We observe that at low energies the model is unable to distinguish between $\gamma$ and $\pi^{0}$, but the accuracy sharply increase with the energy and become a plateau around $\sim 70\%$ accuracy after 900 MeV energy. 
\newpage

\begin{figure}[h!]
    \centering
    \includegraphics[scale=0.6]{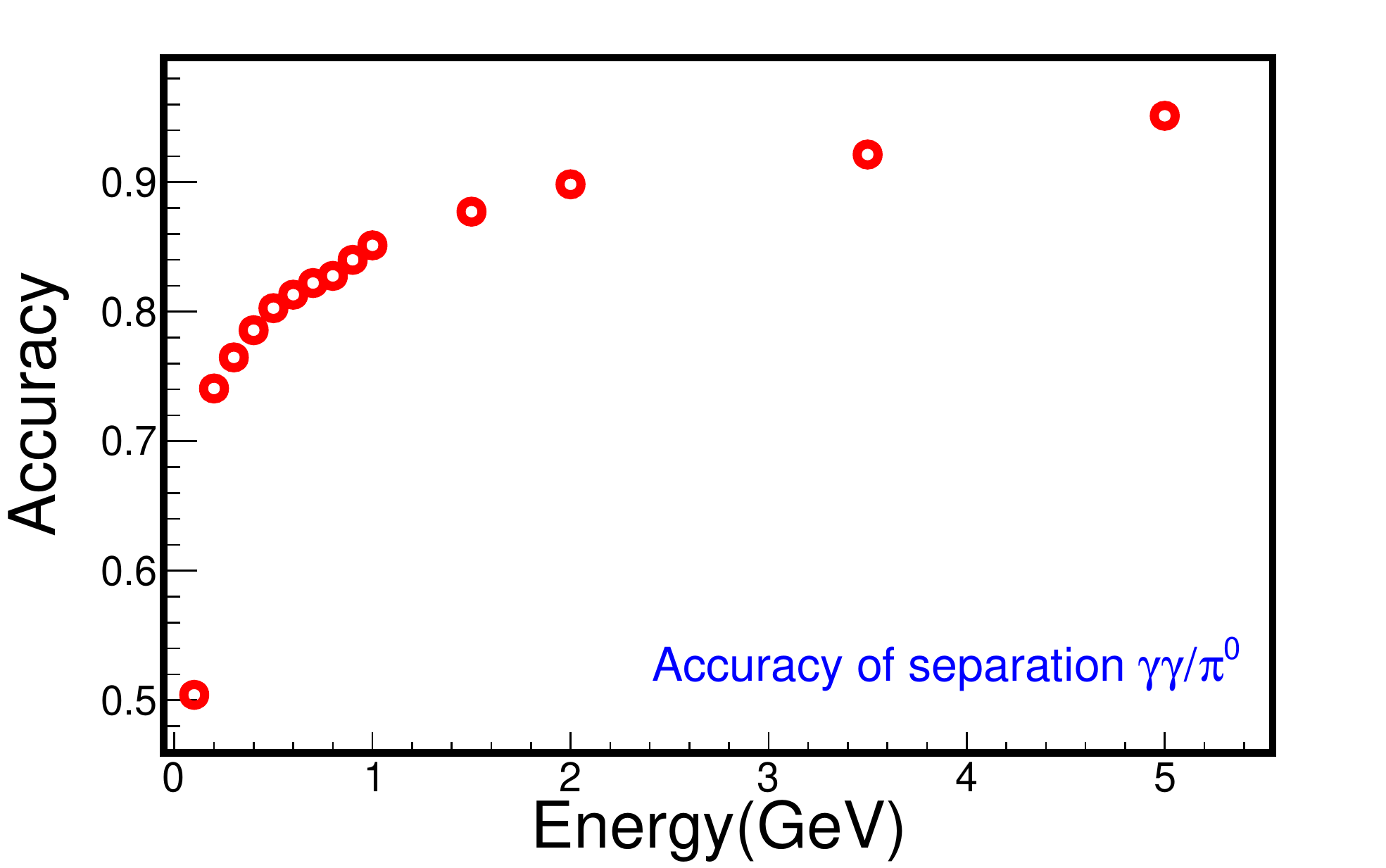}
    \caption{Accuracy of classification between $\gamma\gamma/\pi^{0}$ by 3D ConvNet model  }
    \label{fig:double}
\end{figure}

We extend the application of our deep network model to distinguish between $\gamma\gamma$ vs $\pi^{0}$. As we can see from Fig.~\ref{fig:double}, there is a smooth rise in the accuracy with increase in the energy and performing good at higher energies by giving an accuracy $>90\%$.

\begin{figure}[h!]
    \centering
    \includegraphics[scale=0.6]{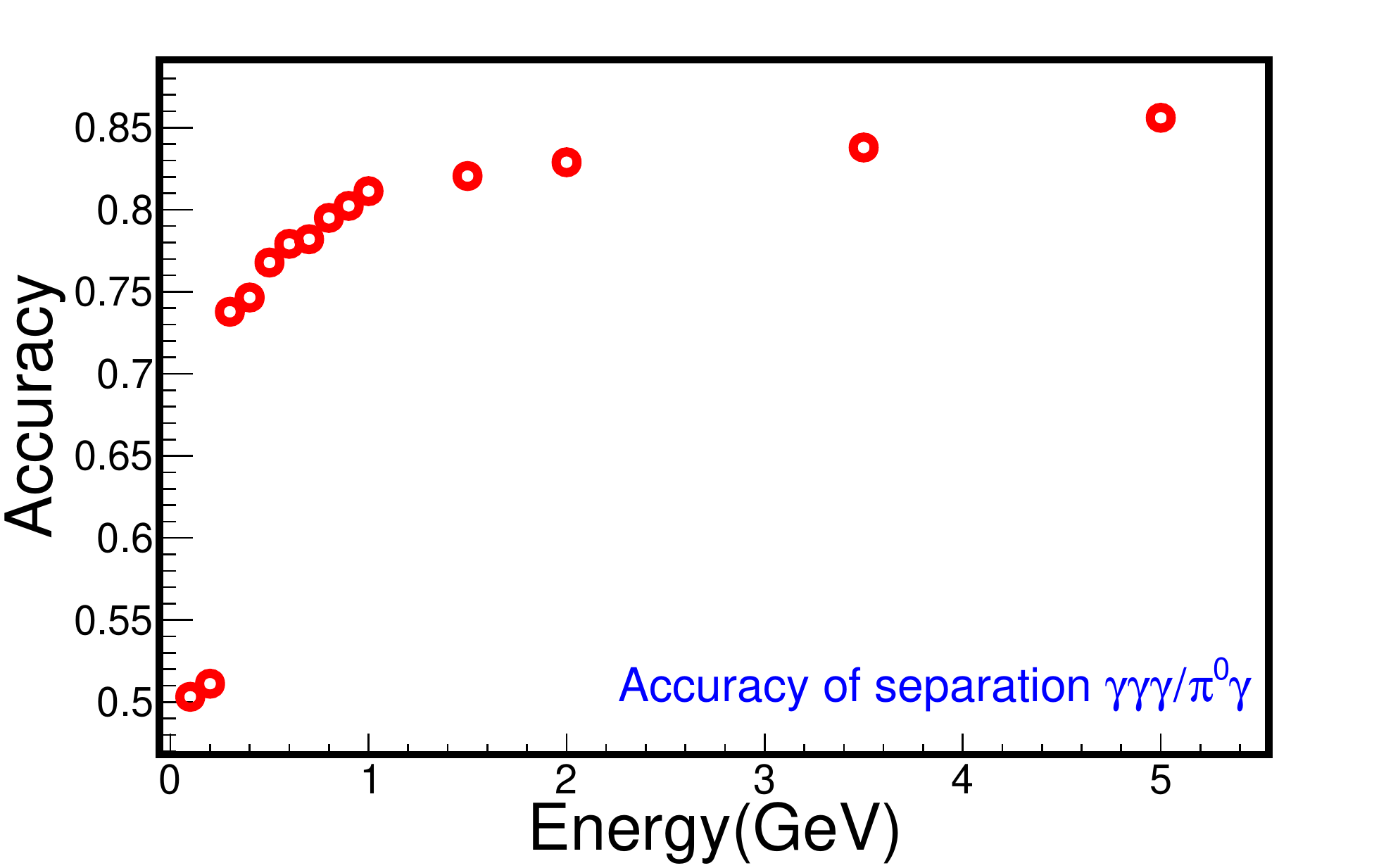}
    \caption{Accuracy of classification between $\gamma\gamma\gamma/\pi^{0}\gamma$ by 3D ConvNet model  }
    \label{fig:triple}
\end{figure}

For the final task, we apply our deep model to classify shower profiles of multi-particle states i.e. $\gamma\gamma\gamma$ vs $\pi^{0}\gamma$. Accuracy vs energy profile for the model is shown in Fig.~\ref{fig:triple}. We infer that the accuracy rises continuously as a function of energy and performs better at higher energies.

\section{Conclusion}
\label{sec:conc}

In this article, we demonstrated the effectiveness of deep learning techniques for the shower identification of single and multi-particle states in the electromagnetic calorimeter. We utilize data from a full detector simulation, based on a complete GEANT4 simulation of electromagnetic calorimeter. The method is based on 3D convolutional neural networks which could outperform traditional and resource consuming techniques.  Our studies, indicate that the model is able to distinguish between showers coming from single as well as multi-particle states from low energy to high energy. In addition, our studies indicate the rigorousness and robustness of the model towards characterizing the shower based on only electromagnetic shower as an input not hadronic shower in hadronic calorimeter or any other input. These results offers a solid direction for implementing essential machine learning based algorithms which are robust to the event by event fluctuations, model-dependent effects present in the data which is a notable application of machine learning at collider experiments.

\acknowledgments
The simulation and training works were carried-out in the computing facility in EHEP Lab at IISER Mohali.



\begin{thebibliography}{99}
\bibitem{geant}
S. Agostinelli et al. GEANT4–a simulation toolkit. Nucl. Instrum. Meth. A,\textbf{ 506},$250-303$, 2003
\bibitem{mrpc}
E. Cerron et al. The Multigap Resistive Plate Chamber Nucl. Instrum. Meth. A,  \textbf{374}, $132-136$,1996
Author, \emph{Title}, \emph{J. Abbrev.} {\bf vol} (year) pg.

\bibitem{cms}
 The CMS Collaboration et al 2008 JINST \textbf{3} S08004
 
\bibitem{alice}
The ALICE Collaboration et al 2008 JINST \textbf{3} S08002

\bibitem{atlas}
 The ATLAS Collaboration et al 2008 JINST \textbf{3} S08003
\bibitem{ml1}
Bruce H. Denby, Neural networks and cellular automata in experimental high-energy physics. Comput. Phys. Commun. \textbf{49}, 429–448(1988)


\bibitem{ml2}
Carsten  Peterson,  Track  finding  with  neural  networks.  Nucl.Instrum.  Methods.  A \textbf{279},  537  (1989)



\bibitem{ml3}
P. Abreu et al., Classification of the hadronic decays of the $Z^{0}$ into b and c  quark  pairs  using  a  neural  network.  Phys.  Lett.  B \textbf{295},383–395 (1992)


\bibitem{ml4}
Dawit Belayneh et al., Calorimetry with deep learning: particle simulation and reconstruction for collider physics, Eur. Phys. J. C (2020) \textbf{80}, 688


\bibitem{ml5}
P. Baldi, K. Bauer, C. Eng, P. Sadowski, D. Whiteson, Jet substructure  classification  in  high-energy  physics  with  deep  neural  networks.  Phys.  Rev.  D  (2016).


\bibitem{ml6}
P. Baldi, P. Sadowski, D. Whiteson, Searching for exotic particlesin high-energy physics with deep learning. Nat. Commun. (2014)


\bibitem{ml7}
Yogesh Verma and Satyajit Jena, Particle Track Reconstruction using Geometric Deep Learning, arxiv:2012.08515

\bibitem{ml8}

L.  M.  Dery,  B.  Nachman,  F.  Rubbo,  A.  Schwartzman,  Weakly supervised  classification  in  high  energy  physics.  J.  Phys.  Conf.Ser. (2018)

\bibitem{pt}
S.~Jena and R.~Gupta, A unified formalism to study transverse momentum spectra in heavy-ion collision,
Phys. Lett. B \textbf{807}, 135551 (2020)

\bibitem{hllhc}
G.  Apollinari,  I.  Béjar  Alonso,  O.  Brüning,  P.  Fessia,  M.  Lam-ont,  L.  Rossi,  L.  Tavian,  High-Luminosity  Large  Hadron  Col-lider (HL-LHC): Technical Design Report V. 0.1. CERN Yellow Reports: Monographs. CERN, Geneva (2017).
\bibitem{pion}
A.~Ayala, J.~Barreiro and L.~M.~Montano, Density and expansion effects on pion spectra in relativistic heavy ion collisions,
Phys. Rev. C \textbf{60}, 014904 (1999)

\bibitem{keras}
Francois  Chollet  et  al.  Keras. \url{https://github.com/keras-team/keras} (2015)

\bibitem{tensorflow}

M. Abadi et al. Tensorflow: Large-scale machine learning on heterogeneous systems. Software available from \url{www.tensorflow.org.} (2015)
\bibitem{hllhc}
G.  Apollinari,  I.  Béjar  Alonso,  O.  Brüning,  P.  Fessia,  M.  Lam-ont,  L.  Rossi,  L.  Tavian,  High-Luminosity  Large  Hadron  Col-lider (HL-LHC): Technical Design Report V. 0.1. CERN YellowReports: Monographs. CERN, Geneva (2017).

\bibitem{reconst}
Sanjib Muhuri et al., Test and characterization of a prototype silicon-tungsten electromagnetic calorimeter, Nucl. Instrum. Meth. A \textbf{764}, $24-29$,2014

\bibitem{reconst2}
M. Chadeeva, "Hadronic shower reconstruction in an imaging calorimeter," 2011 IEEE Nuclear Science Symposium Conference Record, Valencia, Spain, 2011, pp. $2135-2140$

\bibitem{reconst3}
Frédéric Juget 2009 J. Phys.: Conf. Ser. \textbf{160} 012033

\bibitem{reconst}
Sanjib Muhuri et al., Test and characterization of a prototype silicon-tungsten electromagnetic calorimeter, Nucl.Instrum.Meth.A 764 (2014) 24-29


\bibitem{reconst4}
G.S. Bitsadze et al., Reconstruction of the coordinate and energy of the electromagnetic shower in the lead-glass hodoscope calorimeter at different entrance angles of 5 GeV positrons, Nucl. Instrum. Meth. A, \textbf{311}, 1992

\bibitem{cnn}
S. Ji, W. Xu, M. Yang and K. Yu, 3D Convolutional Neural Networks for Human Action Recognition,"in IEEE Transactions on Pattern Analysis and Machine Intelligence, vol. \textbf{35}, no. 1, pp. 221-231, Jan. 2013






\end{thebibliography}
\end{document}